\documentclass[letterpaper,12pt]{article}   
\usepackage{osajnl2} 
\usepackage{amsmath}
\pdfoutput=1

\begin{document}

\title{Finite element modeling of coupled optical microdisk resonators for displacement sensing}

\author{I. S. Grudinin,$^{*}$ and Nan Yu}
\address{$^1$Jet Propulsion Laboratory, California Institute of Technology, \\ 4800 Oak Grove dr., Pasadena, CA 91109, USA}
\address{$^*$Corresponding author: grudinin@jpl.nasa.gov}

\section*{} 
We analyze normal mode splitting in a pair of vertically coupled microdisk resonators. A full vectorial finite element model is used to find the eigen frequencies of the symmetric and antisymmetric composite modes as a function of coupling distance. We find that the coupled microdisks can compete with the best Fabry-Perot resonators in displacement sensing. We also show how we configured FreeFem++ for the sphere eigenvalue problem.

\section{Introduction}
In whispering gallery mode (WGM) resonators the electromagnetic field is trapped inside a circular dielectric by total internal reflections from the boundary\cite{richtmeyer, oraevskyen, book}. Modes of such resonators are distinguished by compact volume and quality factors as high as $Q>10^{11}$ \cite{million} in a broad range of frequencies. Demonstrated applications of such resonators include microwave cavities for atomic clocks \cite{dick}, cavity optomechanics \cite{kipvah}, nonlinear and quantum optics\cite{ilchrev, vahnature,carmonuv}, frequency standards\cite{matsko}, biosensors \cite{armani, arnold, lu}, microcavity frequency combs \cite{grudinincomb}, RF photonics \cite{mhz}, optical clocks and ultrastable lasers \cite{oewaveslaser}. 

When two optical microresonators are placed side by side, the weak evanescent field just outside the surface mediates energy exchange, resulting in optical coupling and normal mode splitting \cite{meter,ultimate, coupledreview}. The coupled microsphere resonators have been shown to have better displacement sensitivity than those based on Fabry--Perot resonators \cite{meter}. This sensitivity results from the strong dependence of the normal mode splitting of the coupled microtoroid resonators on the gap between the cavities \cite{phonon}. A further improvement in sensitivity over the edge--coupled configuration was predicted for vertically coupled hemispherical resonators \cite{meter}. The experimental realization of such hemispheres has remained unattainable. However, recently developed on--chip microdisk resonators \cite{microdisks} can be fabricated with sufficient control over spectra to make vertically coupled resonator configuration possible.

We here study the normal mode splitting of two identical vertically coupled microdisk resonators. Using the finite element method (FEM) we find normal mode splitting (NMS) for idealized coupled resonator geometries. We also provide a basic noise performance analysis of such displacement transducer.

\section{WGM eigenvalue problem with FEM}
Analytical solutions for WG modes are only available for a limited number of ideal geometries, such as a sphere or an ellipsoid \cite{richtmeyer,oraevskyen,spheroid,book}. The three dimensional vectorial Maxwell wave equation can be solved numerically, taking the axial symmetry of WGM resonators into account. In this approach \cite{webb, oxborrow}, the problem can be reduced to a system of 3 coupled equations in a cylindrical 2D ``$\rho-z$'' section of a resonator. The spurious modes resulting from the improper accounting of the curl operator zero space \cite{wong} are partially suppressed by using the penalty coefficient method, enforcing the $\nabla\cdot H=0$ condition approximately.

We used a scriptable and scalable FreeFem++ solver \cite{freefem}, which utilizes ARPACK\cite{arpack} for eigenvalue problem solutions. We used UMFPACK\cite{umfpack} as a matrix solver due to its better handling of large matrices compared to other solvers available in FreeFem. We here show \cite{freecode} how to set a simple spherical eigenvalue WGM problem in FreeFem++ following the approach of ref. \cite{oxborrow}. The results obtained with FreeFem++ are very close to those obtained with the commercial package COMSOL Multiphysics, taking into account the lack of easy mesh portability between the two.

\section{A dielectric sphere}
We start with the weak formulation \cite{oxborrow} to compute the fundamental TE and TM mode frequencies and field distributions in a fused silica sphere with refractive index of 1.46 and azimuthal index $M=10^3$. Exact analytical complex eigen values can be obtained by numerically solving the following equation \cite{book}
\begin{equation}\label{precise}
y\left(Pnh_M^{(1)}(y)j_{M-1}(ny)-j_M(ny)h_{M-1}^{(1)}(y)\right)+M(1-P)(h_M^{(1)}(y)j_M(ny))=0
\end{equation}
Here $j_M(ny)$ and $h_M^{(1)}(y)$ are the spherical Bessel and Hankel functions of order $M$, $y\equiv ka$, $a$ is resonator's radius, $n$ is the refractive index. $P=1$ for TE modes and $P=1/n^2$ for TM modes.  The imaginary parts of the numerical solutions give radiative quality factor of the mode: $Q_{rad}=\frac{Re(y)}{2Im(y)}$. Simplified real equation may also be solved, where Hankel functions are replaced with the Neumann function.

\begin{figure}[htbp]
\centering
\includegraphics[width=7cm]{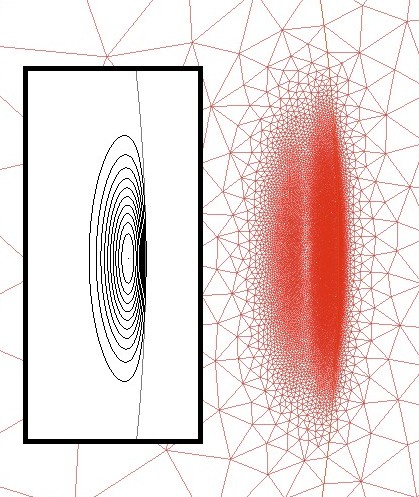}
\caption{Adapted mesh in mode localization region. Inset: FEM solution for a sphere -- $TM_{1000,1000,1}$ mode. Iso-value lines are equally spaced, scale for the mesh and the solution is the same.}
\label{meshes}
  \end{figure}

We can estimate the error of the FEM solutions as $\delta y=|(y_{fem}-y_{exact})|$. The TE and TM solutions are generated on meshes that were adapted to the magnitude of magnetic field vector. The mesh and the mode profile of a TM mode are shown in Fig. \ref{meshes}.
 The results are shown in Table 1. The absolute frequency error is given by $\delta F=\delta y (c/2\pi a)$. For our mesh this corresponds to around 80 MHz (TE) for a sphere with radius $a=36\mu m$.
\begin{table}[htbp]
{\bf \caption{Computed errors of $y=ka$ for a sphere$^a$, $M=10^3, n=1.46$}}
\begin{center}
\begin{tabular}{ccc} \hline
 & TE & TM \\ \hline
mesh size$^b$ &34120/17072 & 37466/18745 \\
$\delta y$ & $6.2\times 10^{-5}$& $7.6\times10^{-6}$\\
\hline
\end{tabular}
\end{center}
\footnotesize $^a$ Second order Lagrange triangular finite elements were used. Boundary conditions is electric wall on the computational window edges. Sphere radius is $a=36\mu m$, computation window height is $20\mu m$, left edge is at $a-8 \mu m$, right edge is at $a+5\mu m$. $^b$ number of triangles / number of vertices.  \normalsize
\end{table}
It shows that the FEM formulation used here, as implemented in FreeFem++, provides good precision for eigen modes of a sphere.  In small microspheres and non-spherical resonators the modes are no longer pure TM and TE, but have hybridized polarizations. Since the weak formulation used here assumes no approximations and models the full field vector, it can be used for arbitrary axially symmetric resonator geometries and expected to retain good precision.

\section{Vertically coupled microdisk resonators}
\begin{figure}[htbp]
\centering
\includegraphics[width=16cm]{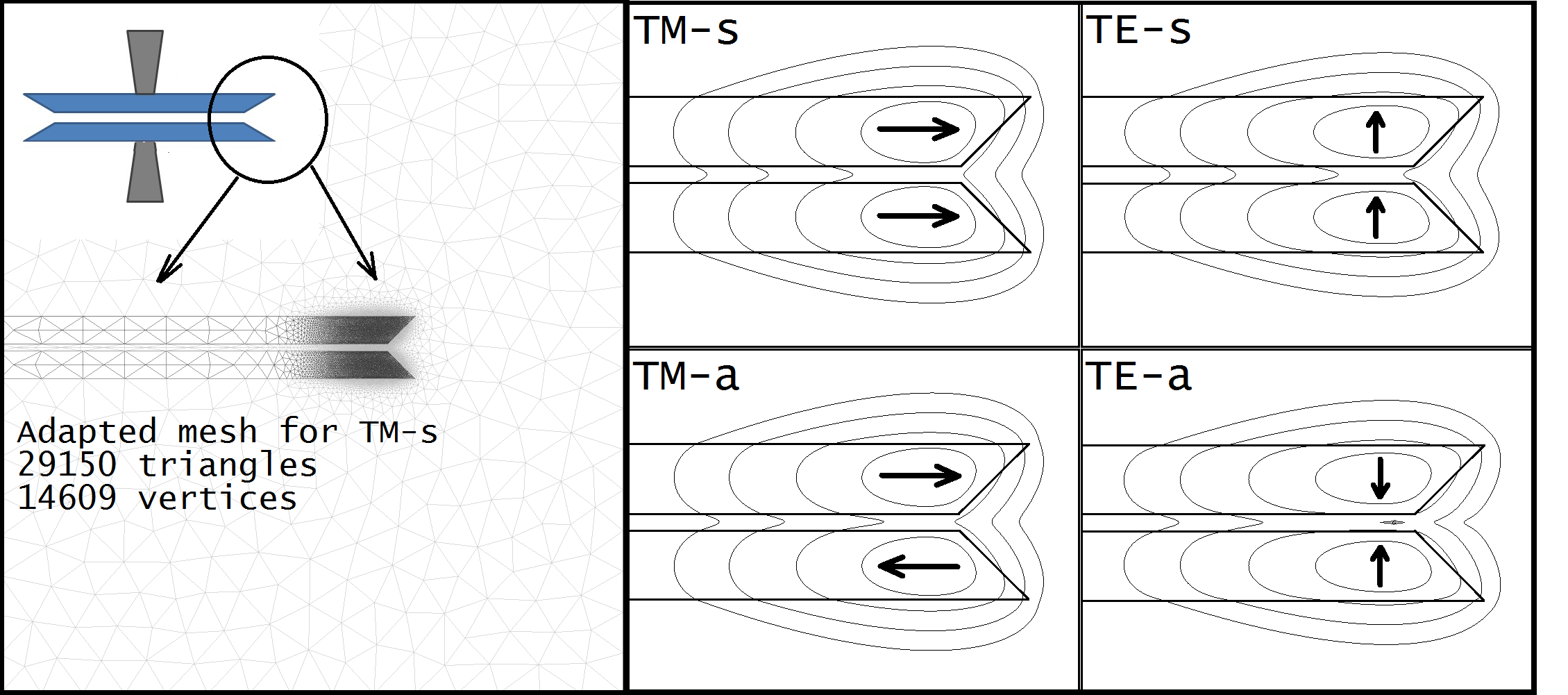}
\caption{Vertical coupling of two silica microdisks with refractive index $n=1.46$, thickness 2 $\mu m$, wedge angle $45^\circ$, radius 180 $\mu m$, air gap 0.5 $\mu m$, and the mode orbital index $M=588$ corresponding to optical wavelength of about 1.56 $\mu m$. The amplitude of the magnetic field vector is shown with lines of equal value. From outside, the lines are $H_{max}/1000, H_{max}/100, H_{max}/10$, and $H_{max}/2$. Approximate electric field vector directions are shown with arrows. }
\label{fig:geometry}
\end{figure}

We used the FEM tool to compute eigen modes of a pair of vertically coupled microdisk resonators. The computation window with typical resonator geometry is shown in Fig. \ref{fig:geometry}. The figure also shows symmetric (s) and antisymmetric (a) composite modes for TE and TM polarizations \cite{coupledreview}. 
 The modes found by the solver were classified as TE or TM by comparing $E^2_z$ and $E^2_r$ integrated over the computational window. For TE modes most of the energy resides in the $E_z$ component. 

\begin{figure}[htbp]
\centering
\includegraphics[width=8cm]{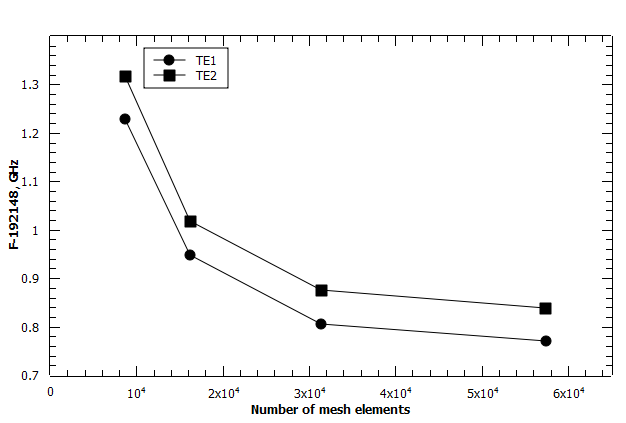}
\caption{Convergence of TE composite mode frequencies with increasing number of mesh elements. Vertical axis shows optical frequency offset by 192 THz. Air gap 2.5 $\mu m$, R=1750 $\mu m$, M=$10^4$.}
\label{fig:conv}
\end{figure}

\begin{figure}[htbp]
\centering
\includegraphics[width=12cm]{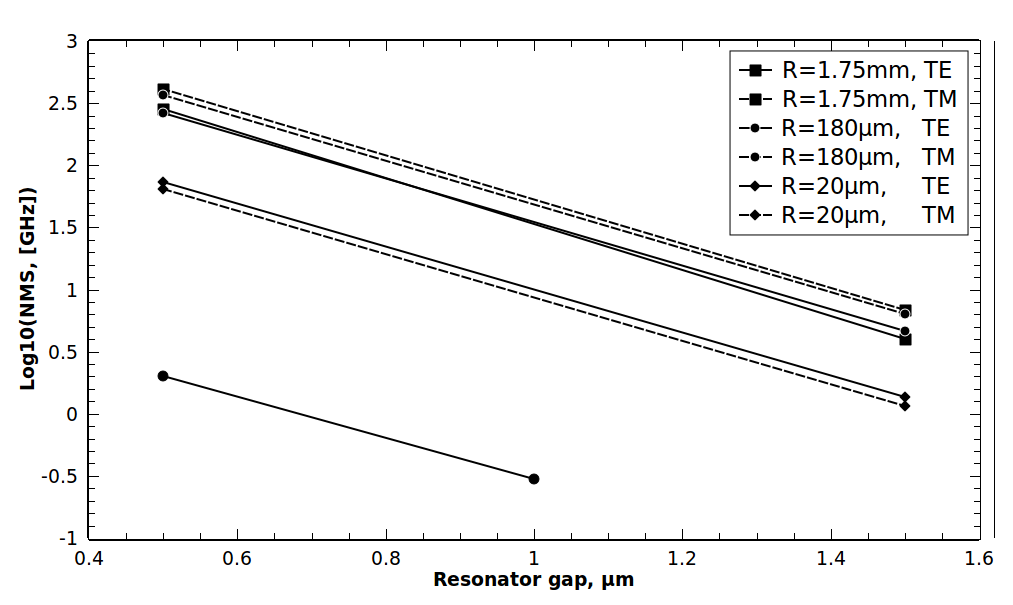}
\caption{Mode splitting in vertically coupled microdisks. NMS of silica microtoroids coupled side--by--side is shown for comparison (adapted from \cite{phonon}). Error bars are similar in size to point markers and are not shown.}
\label{fig:nms}
\end{figure}

We investigated the dependence of NMS on the air gap for modes with the wavelength around 1550 nm, with disk radii of 20 ($M=10^2$), 180 ($M=10^3$) and 1750 ($M=10^4$) micrometers. For each mode, we started with a homogeneous mesh with approximately $10^4$ triangular elements. The mesh was then optimized to the magnetic field vector amplitude of each mode and the new frequency was found. The number of mesh elements was then doubled until reasonable converged solution is achieved. The computation error was chosen as the difference between frequency obtained with the final mesh and that obtained with the previous mesh having half the elements. The mesh refinement was repeated until computation error was less than 0.5 NMS. We observed good convergence of our solutions, as shown on Fig. \ref{fig:conv}.
The computation window edges were also appropriately spaced: the change in frequency due to offset of the walls outwards by 1 micrometer was much smaller than the computation error of the mode. Since the dependence of NMS on air gap is exponential as expected \cite{meter}, it was sufficient to only find NMS for two values of the air gap. We adjusted the air gap, keeping the edges of the computational window unchanged. The results are presented in Fig. \ref{fig:nms}.
From the data we obtain the dependence for NMS, $S=10^{kx+b}$ [GHz] as a function of the gap $x$ [$\mu $m] for TM modes, as shown in Table \ref{tabexp}. The effective gap \cite{meter} characterizing tunability at optical frequency $\nu\simeq193$ THz is given by $d=\left( \frac{1}{\nu} \frac{\partial S}{\partial x}\right)^{-1}=\left( \frac{1}{\nu} ln(10)kS\right)^{-1}$.
\begin{table}[htbp]
{\bf \caption{Displacement sensitivities of coupled microdisks for optical frequency around 193 THz (wavelength $1.56 \mu m$). Typical quality factor is $Q\simeq5\times10^8$ for microdisks and $Q\simeq\times10^7$ for coupled toroids \cite{phonon}. The LIGO and the capacitive sensor are listed for reference. }\label{tabexp}} 
\begin{center}
\begin{tabular}{cccccc} \hline
diameter, $\mu m$ & M & k, $\mu m^{-1}$ & b & d, cm (at x=0.5$\mu m$)& d/Q, cm\\ \hline
20& $10^2$& -1.744 & 2.682 &0.075 & $1.5\times10^{-10}$ \\
180 & $10^3$ & -1.763 & 3.45 & 0.013 &$2.6\times10^{-11}$\\
1750 & $10^4$ & -1.779 & 3.507& 0.011 &$2.2\times10^{-11}$  \\
66 & Toroids & -1.647 & 1.125& 2.45 & $2.45\times10^{-7}$\\
 Advanced LIGO \cite{ligo} & & & &  &$1.2\times10^{-9}$ \\
 Capacitive sensor \cite{capacitive} & & & & & $6\times10^{-9}$ \\
 \hline
\end{tabular}
\end{center}
\end{table}

One observation from Fig. \ref{fig:nms} is that compared to silica microtoroids coupled side-by-side \cite{phonon}, the NMS in vertically coupled microdisks is significantly larger. Another interesting result is that the mode splitting saturates for large disks: NMS is virtually the same for the disks with radius of 1750 $\mu m$ as it is for the disks with radius of 180 $\mu m$. This may be explained by the increased mode confinement in vertical direction in larger disks. The increased confinement leads to smaller field overlap and coupling. In addition, once the length of the guided whispering gallery mode exceeds the geometry-specific interaction length\cite{coupledreview} the maximum energy exchange has been achieved and no further increase of coupling constant is possible. Indeed, the TM-mode NMS for $M=1.5\times 10^4$, $R=2620 \mu m$ is only 0.2\% larger than for $M=10^4$, $R=1750 \mu m$. This indicates that further optimization of disk radius and thickness is possible, which is beyond the scope of this paper. From physical considerations it was expected that vertically coupled configurations will be capable of operating at larger air gaps compared to edge-coupled resonators. Indeed we find that that vertically coupled configuration provides NMS similar to coupled microtoroids at roughly double the gap distance. Finally, we observe that while for smaller disks TE modes have larger NMS, for larger disks the TM modes become split more than TE. This is explained by the influence of the disk wedge. The decrease of the wedge angle is also known to push the mode closer to the disk center, thus we assume that to first order this corresponds to decreasing the disk's radius.

\section{Displacement measurement}
 For a realistic optical quality factor of $Q=10^8$ and a pair of resonators with $M=10^4$ TM modes operating at 0.5 $\mu m$ gap we derive the minimum measurable displacement \cite{meter}:  
 \begin{equation}
 \Delta x=\frac{d}{Q}\sqrt{\frac{h\nu}{W\tau}}=2.2\times10^{-11}\sqrt{\frac{h\nu}{W\tau}}
 \end{equation}
Here W is the laser pump power. For example, for $W=100 \mu W$, and the averaging time of 1 s the minimum displacement is $8\times10^{-18}$ cm. This sensitivity is comparable or better than what can be achieved with state of the art Fabry--Perot, and superconductive microwave cavities, however the on-chip microdisk resonators have small mass of moving parts and can be implemented in a compact measurement setup. For comparison, the effective coefficient of a Fabry--Perot cavity is given by finesse: $d/Q=\lambda/2F$. For state of the art resonators $F\simeq 2\times 10^6$  \cite{Rempe}, giving $d/Q\simeq1.5\times10^{-11}$cm.

There are a number of factors that can limit the practical sensitivity of the coupled resonator sensor. In addition to photon shot noise, there are such fundamental factors as Brownian and thermorefractive\cite{gorgru, matsko} noises.  One should also avoid the threshold of Kerr and thermal nonlinearity. However, they are specific to a particular NMS measurement scheme and resonator geometries. No nonlinearity was observed for some of the larger microdisks for up to 1 mW  of pump power \cite{microdisks}. Thermorefractive noise will be smaller in larger cavities. It was also experimentally found that the Q factor noticeably degraded for the smallest gaps for the coupled toroids \cite{phonon}. Similarly, it can be expected that the quality factor will depend on the air gap for the vertically coupled disks. It is possible to improve the FEM code to directly compute the radiative Q factor of this system, as will be reported elsewhere. 

The ultimate sensitivity of the vertically coupled sensor will depend on the way the NMS is measured. In a passive scheme, laser noise and mechanical noises will determine the sensitivity of the system. One alternative approach is to use an active sensor configuration, where, for example, one of the resonators is doped to provide optical gain. Upon pumping the coupled resonators, lasing into normal symmetric and antisymmetric modes will occur (e.g. similar to ref. \cite{fluorescent}) . The beatnote of the two can be detected with a photodetector, directly providing the NMS. The sensitivity will be limited by the beatone linewidth, which is in turn limited by the lasing linewidth. While erbium fiber lasers have linewidth less than 10 KHz \cite{erlaser}, alternative active schemes may also be used. Coupled resonators will experience attractive and repulsive forces \cite{povinelli}, the magnitude of which can be directly estimated from the tuning curves. 

\section{Conclusion}
We have presented a finite element analysis of the normal mode splitting in a system of two vertically coupled silica microdisks. The analysis utilizes a publicly available FEM solver FreeFem++. We find that with the recent advances in microdisk fabrication \cite{microdisks} the vertically coupled resonators have more than two orders of magnitude better displacement sensitivity compared to edge-coupled system. In addition, the sensitivity gain saturates for larger disk diameters. The ultimate displacement measurement sensitivity of this system is comparable to a Fabry-Perot system, while having a much smaller size. The vertically coupled configuration opens new opportunities for scientific and industrial applications of displacement measurements.

\section{Acknowledgments} The research described in this paper was carried out at the Jet Propulsion Laboratory,
California Institute of Technology, under a contract with the NASA. We are grateful to Professors K. J. Vahala, M. L. Gorodetsky, F. Hecht and participants of FreeFem mailing list for helpful discussions.

\end{document}